%% file: main.tex
\documentclass[letterpaper]{IEEEtran}
\usepackage{caption,graphicx,cite}
\usepackage{amsmath,amssymb,mathtools,bm,etoolbox}
\newcommand{\e}[1]{{\mathbb E}\left[ #1 \right]}

\usepackage{stackengine}
\usepackage{cuted}

\usepackage{lipsum}
\usepackage{suffix,float}
\DeclarePairedDelimiterX\MeijerM[3]{\lparen}{\rparen}%
{\begin{smallmatrix}#1 \\ #2\end{smallmatrix}\delimsize\vert\,#3}
\newcommand\MeijerG[8][]{%
  G^{\,#2,#3}_{#4,#5}\MeijerM[#1]{#6}{#7}{#8}}
\WithSuffix\newcommand\MeijerG*[7]{%
  G^{\,#1,#2}_{#3,#4}\MeijerM*{#5}{#6}{#7}}
\newcommand\ddfrac[2]{\frac{\displaystyle #1}{\displaystyle #2}}
\usepackage{amsthm}
\usepackage{subcaption}
\newtheorem{theorem}{Theorem}
\newtheorem{corollary}{Corollary}

\begin{document}

\title{Temporal CSI Correlation in Mixed RF/FSO Cooperative Relaying Systems Under Joint Effects of HPA Nonlinearities and IQ Imbalance}


\author{Elyes~Balti,~\IEEEmembership{Member,~IEEE,} Neji~Mensi,~\IEEEmembership{Member,~IEEE,} and Danda~B.~Rawat,~\IEEEmembership{Senior Member,~IEEE}
\thanks{Elyes Balti is with the Wireless Networking and Communications Group, Department of Electrical and Computer Engineering, The University
of Texas at Austin, Austin, TX 78712 USA e-mail: ebalti@utexas.edu.}
\thanks{Neji Mensi and Danda B. Rawat are with the Department
of Electrical Engineering and Computer Science, Howard University, Washington, DC, 20059 USA e-mail: neji.mensi@bison.howard.edu, danda.rawat@howard.edu.}}

\maketitle

\begin{abstract}
In this paper, we present the performance analysis of mixed RF/FSO system with multiple relays. To select the best relay, we adopt partial relay selection with outdated CSI wherein we investigate the effect of the temporal correlation of the channels. Unlike the vast majority of work, we introduce the impairments to the relays and the destination and we compare the performance against conventional RF relaying systems. We further derive the expressions of the outage probability and the ergodic capacity as well as the bounds to unpack engineering insights into the system robustness.
\end{abstract}


\IEEEpeerreviewmaketitle

\input{introduction}
\input{model}

\input{outage}

\input{capacity}
\input{numerical}

\input{conclusion}

\bibliographystyle{IEEEtran}
\bibliography{main}
\end{document}

%% file: introduction.tex
\section{Introduction}
Wireless optical communications also known as Free-Space Optic (FSO) is considered as the key stone for the next generation of wireless communication since it has recently gained enormous attention for the vast majority of the most well-known networking applications such as fiber backup, disaster recoveries and redundant links \cite{uysal}. The main advantages of employing the FSO is to reduce the power consumption and provide higher bandwidth. Moreover, FSO becomes as an alternative or a complementary to the RF communication as it overcomes the problems of the spectrum scarcity and its license access to free frequency band. In this context, many previous attempts have leveraged some these advantages by introducing the FSO into classical systems to be called Mixed Radio-Frequency (RF)/FSO systems \cite{soleiman,joint,tractable,j2,cochannel,c2,thesis}. This new system architecture reduces not only the interference level but also it offers full duplex Gigabit Ethernet throughput and high network security \cite{mmwaves,fd1,neji1,adaptive}. Although the literature has shown the superiority of the mixed RF/FSO systems over the classical RF systems, they still suffer from the reliability scarcity and power efficient coverage. To overtake this difficulty, previous research attempts have proposed cooperative relaying techniques hybridized with the mixed RF/FSO systems since it improves not only the capacity of the wireless system but also it offers high Quality of Service (QoS) \cite{sub6,yassine1,yassine2,neji3}. Recently, this new efficient system model has attracted considerable attention in particular using various relaying schemes. The most common used relaying techniques are Quantize-and-Encode \cite{quantizeencode}, Decode-and-Forward \cite{tractable,stochastic} and Amplify-and-Forward \cite{neji2,c1,c3}. In practice, however, the hardware (source, relays, destination) are susceptible to impairments, e.g., High Power Amplifier (HPA) nonlinearities \cite{j1,17} phase noise \cite{19} and In Phase and Quadrature (IQ) imbalance \cite{20}. Due to its low quality and price, the relay suffers from the nonlinear Power Amplifier (PA) impairment which is caused primarily by the non-linear amplification of the signal that may cause a distortion and a phase rotation of the signal. The most common nonlinear HPA model are Traveling Wave Tube Amplifier (TWTA), Soft Envelope Limiter (SEL) \cite{22} and Ideal Soft Limiter Amplifier (ISLA) \cite{23}. Maletic \textit{et al.} \cite{22} concluded that the SEL has less severe impact on the system performance than the TWTA model. Furthermore, there are few attempts \cite{2,asymmetric} considering mixed RF/FSO system affected by a general model of impairments but they did not specify the type/nature of the hardware impairments. In this work, we propose a mixed RF/FSO system with multiple relays employing Fixed Gain (FG) relaying. Since the channels are subject to time selectivity due to the mobility and Doppler spread, we assume partial relay selection (PRS) based on the outdated Channel State Information (CSI) of the first hop \cite{2}. Besides, we assume that the relays are vulnerable to either SEL or TWTA impairments while the destination suffers from IQ imbalance. The rest of this paper is organized as follows: the system model is presented in Section II. The analysis of the outage probability and the ergodic capacity is provided in Sections III and IV, respectively. Section V discusses the numerical and simulation results while the concluding remarks and the future directions are given in Section VI.

%% file: model.tex
\section{System Model}
\begin{figure}[t]
    \centering
    \includegraphics{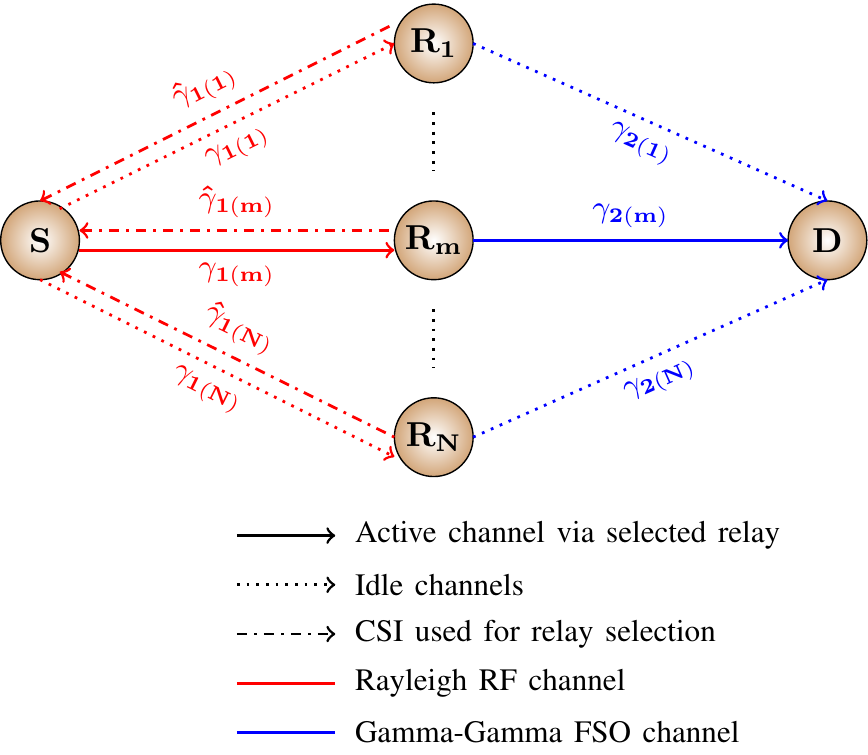}
    \caption{Mixed RF/FSO system with partial relay selection}
    \label{system}
\end{figure}

The system consists of a source ($S$) communicating with a destination ($D$) though $N$ parallel relays shown by Fig.~\ref{system}. For a given transmission, $S$ periodically receives the CSIs ($\gamma_{1(\ell)}$ for $\ell = 1\ldots \textit{N})$ of the first hop from the \textit{N} relays and sorts them in an increasing order of magnitude as follows: $\gamma_{1(1)}\leq\gamma_{1(2)}\leq \ldots \leq\gamma_{1(N)}$. The perfect scenario is to select the best relay \textit{(m = N)} but this best one is not always available. In this case, $S$ will select the next best available relay. Consequently, the PRS protocol selects the \textit{m}-th worst or $(N-m)$-th best relay $R_{m}$. Given that the feedback is delayed due to the time selectivity caused by the mobility and Doppler spread, the CSI at the time of selection is different from the CSI at the instant of transmission. In this case, outdated CSI should be assumed instead of perfect CSI estimation. Hence, the instantaneous CSI used for relay selection $\tilde{\gamma}_{1(m)}$ and the instantaneous CSI $\gamma_{1(m)}$ used for transmission are correlated with the temporal correlation $\rho$.
The received signal at the $m$-th relay is given by
\begin{equation}
y_{1(m)} = h_m s + \nu_1
\end{equation}
where $s \in \mathbb{C}$ is the information signal, $h_m$ is the RF fading between $S$ and $R_{m}$ and $\nu_1$  $\backsim$ $\mathcal{CN}$ (0, $\sigma^2_0$) is the Additive White Gaussian Noise (AWGN).
\subsection{High Power Amplifier nonlinearities at the relays}
The PA nonlinearities impairment is introduced to the relays. The amplification of the signal happens in two time slots. In the first slot, the received signal at the relay $R_{m}$ is amplified by a proper gain $G$ as ${\phi_m} = G y_{1(m)}$. The gain $G$ can be defined as
\begin{equation}
G = \sqrt{\frac{\sigma^2}{\e{|h_m|^2} P_1 + \sigma^2_0}}
\end{equation}
where $\e{\cdot}$ is the expectation operator, $P_1$ is the average transmitted power from $S$ and $\sigma^2$ is the mean power of the signal at the output of the relay block. In the second time slot, the signal passes through a nonlinear circuit $\psi_m = f(\phi_m)$.\\
The PA of the relay is assumed to be memoryless. A memoryless PA is characterized by both Amplitude to Amplitude (AM/AM) and Amplitude to Phase (AM/PM) characteristics. The functions AM/AM and AM/PM transform the signal distortion respectively to $A_m(|\phi_m|)$ and $A_p(|\phi_m|)$ and then the output signal of the nonlinear PA circuit is given by
\begin{equation}
\psi_m = A_m(|\phi_m|)~e^{j(\text{angle}(\phi_m)+A_p(|\phi_m|))}  
\end{equation}
The characteristic functions of the SEL and TWTA impairments models are respectively given by \cite{joint}. From a given saturation level $A_{\text{sat}}$, the relay's PA operates at an input back-off (IBO), which is defined by IBO = $\frac{A^2_{\text{sat}}}{\sigma^2}$. According to Bussgang Linearization Theory \cite{17}, the output of the nonlinear PA circuit linearly depends on both the linear scale $\delta$ of the input signal and a nonlinear distortion $d$ which is uncorrelated with the input signal and follows the circularly complex Gaussian random variable $d \backsim \mathcal{CN} (0,~\sigma^2_d)$. Then, the AM/AM characteristic $A_m(|\phi_m|)$ can be expressed as follows
\begin{equation}
A_m(|\phi_m|) = \delta~|\phi_m| + d.    
\end{equation}
Note that $\delta$ and $\sigma^2_d$ for SEL and TWTA are given by \cite{joint}. Then at the relay $R_{m}$, the RF amplified signal is converted to an optical one which is given by \cite{2}
\begin{equation}
r_{m} = G(1 + \eta \psi_m)
\end{equation}
where $\eta$ is the electrical-to-optical conversion coefficient.
\subsection{In Phase and Quadrature Imbalance at the destination}
In case of perfect IQ mismatch, the received signal at the destination can be expressed as follows
\begin{equation}
y_{2(m)} = I_m G \eta \psi_m + \nu_2
\end{equation}
where $I_m$ is the optical irradiance between the relay $R_{m}$ and $D$, $\eta$ is the optical-to-electrical conversion coefficient, and $\nu_2$  $\backsim$ $\mathcal{CN}$ (0, $\sigma^2_0$) is the AWGN.\\
Given that the destination is affected by IQ imbalance, the received signal is given by
\begin{equation}
\hat{y}_{2(m)} = \omega_1 y_{2(m)} + \omega_2 y_{2(m)}^*
\end{equation}
where $y_{2(m)}^*$ is called the mirror signal introduced by the IQ imbalance at $D$ and the coefficients $\omega_1$ and $\omega_2$ are respectively given by
\begin{equation}
\omega_1 = \frac{1 + \zeta e^{-j\theta}}{2}
\end{equation}
\begin{equation}
\omega_2 = \frac{1 - \zeta e^{j\theta}}{2}
\end{equation}
where $\theta$ and $\zeta$ are respectively the phase and the magnitude imbalance. This impairment is modeled by the Image-Leakage Ratio ({\scriptsize{\textsf{ILR}}}), which is given by ${\scriptsize{\textsf{ILR}}} = \left|\frac{\omega_2}{\omega_1}\right|^2$.\\
For an ideal $D$, $\theta = 0, \zeta = 1, \omega_1 = 1, \omega_2 = 0$, and {\scriptsize{\textsf{ILR}}} = 0.
\subsection{Channels Models}
Since the RF channels are subject to correlated Rayleigh fading, the Cumulative Distribution Function (CDF) of the instantaneous RF {\scriptsize{\textsf{SNR}}} $\gamma_{1(m)}$ is given by \cite[Eq.~(9)]{2}
\begin{equation}
\begin{split}
F_{\gamma_{1(m)}}(x) =& 1 - m{N \choose m}\sum_{n=0}^{m-1} {m-1 \choose n}  \frac{(-1)^n}{N-m+n+1}\\&
\times\exp\left(-\frac{ (N-m+n+1)x }{((N-m+n)(1-\rho)+1)\overline{\gamma}_1}\right).
\end{split}
\end{equation}
Since the instantaneous {\scriptsize{\textsf{SNR}}} $\gamma_{2(m)}$ experiences Gamma-Gamma fading, the Probability Density Function (PDF) is given by
\begin{equation}
f_{\gamma_{2(m)}}(x) = \frac{(\alpha\beta)^{\frac{\alpha+\beta}{2}} x^{\frac{\alpha+\beta }{4}-1}}{2\Gamma(\alpha)\Gamma(\beta)\overline{\gamma}_2^{\frac{\alpha+\beta}{4}}} \MeijerG[\Bigg]{2}{0}{0}{2}{-}{\frac{\alpha-\beta}{2},~\frac{\beta-\alpha}{2}}{\alpha\beta\sqrt{\frac{x}{\overline{\gamma}_2}}}
\end{equation}
where $\MeijerG[\Bigg]{m}{n}{p}{q}{\bold{a}_n,~\bold{a}_p}{\bold{b}_m,~\bold{b}_q}{\cdot}$ is the Meijer-G function, $\alpha$ and $\beta$ are respectively the small-scale and large-scale of the scattering process in the atmospheric environment. These parameters are given by
\begin{equation}
\alpha = \left( \exp\left[ \frac{0.49 \sigma_R^2}{(1+1.11\sigma_R^{\frac{12}{5}})^{\frac{7}{6}}}\right] -1\right)^{-1}
\end{equation}
\begin{equation}
\beta = \left( \exp\left[ \frac{0.51 \sigma_R^2}{(1+0.69\sigma_R^{\frac{12}{5}})^{\frac{5}{6}}}\right] -1\right)^{-1}
\end{equation}
where $\sigma_R^2$ is called Rytov variance which is a metric of the atmospheric turbulence intensity.
\subsection{End-to-end signal-to-noise-plus-distortion ratio (SNDR)}
The average {\scriptsize{\textsf{SNR}}} of the first hop is given by
\begin{equation}
\overline{\gamma}_1 = \frac{P_1|h_m|^2}{\sigma_0^2}    
\end{equation}
While the average {\scriptsize{\textsf{SNR}}} $\overline{\gamma}_2$\footnote[1]{The average {\scriptsize{\textsf{SNR}}} $\overline{\gamma}_2$ is defined as $\overline{\gamma}_2 = \eta^2\e{I_{m}^2}/\sigma_{0}^2$, while the average electrical {\scriptsize{\textsf{SNR}}} $\mu_2$ is given by $\mu_2 = \eta^2\e{I_{m}}^2/\sigma_{0}^2$. Therefore, the relation between the average {\scriptsize{\textsf{SNR}}} and the average electrical {\scriptsize{\textsf{SNR}}} is trivial given that $ \frac{\e{I^2_{m}}}{\e{I_{m}}^2} = \sigma^2_{\text{si}} + 1$, where $\sigma^2_{\text{si}}$ is the scintillation index \cite{scin}.} of the second hop can be expressed as
\begin{align}
 \overline{\gamma}_2 = \frac{\e{I^2_{m}}}{\e{I_{m}}^2}\mu_2   
\end{align}
where $\mu_2$ is the average electrical {\scriptsize{\textsf{SNR}}} given by
\begin{equation}
    \mu_2 = \frac{\eta^2\e{I_{m}}^2}{\sigma_{0}^2}.
\end{equation}
According to \cite[Eq.~(16)]{22}, the end-to-end {\scriptsize{\textsf{SNDR}}} is given by
\begin{equation}
{\scriptsize{\textsf{SNDR}}} = \frac{\gamma_{1(m)}\gamma_{2(m)}}{{\scriptsize{\textsf{ILR}}}\gamma_{1(m)}\gamma_{2(m)}+(1+{\scriptsize{\textsf{ILR}}})(\e{\gamma_{1(m)}} + \kappa\gamma_{2(m)} + \kappa)}    
\end{equation}
where $\e{\gamma_{1(m)}}$ is given by \cite[Eq.~(10)]{2} and the term $\kappa$ is defined as the ratio between the received {\scriptsize{\textsf{SNR}}} and the average transmitted {\scriptsize{\textsf{SNDR}}} at the relay which is given by
\begin{equation}
\kappa = 1 + \frac{\sigma_d^2}{\delta^2G^2\sigma_0^2}.
\end{equation}

%% file: outage.tex
\section{Outage Probability}
\subsection{Exact Analysis}
The outage probability is defined as the probability that the {\scriptsize{\textsf{SNDR}}} falls below a given outage threshold $x$. It can be written as follows
\begin{equation}
P_{\text{out}}({\scriptsize{\textsf{SNDR}}},x) = \mathbb{P}[x \leq {\scriptsize{\textsf{SNDR}}}] = F_{{\scriptsize{\textsf{SNDR}}}}(x)
\end{equation}
where $F_{{\scriptsize{\textsf{SNDR}}}}(\cdot)$ is the CDF of the {\scriptsize{\textsf{SNDR}}}.
\begin{theorem}
Under joint effects of HPA nonlinearities and IQ imbalance, the outage probability is given by (\ref{outage}) if $x<\frac{1}{\text{ILR}}$, otherwise, it is equal to 1.
\end{theorem}
\begin{equation}\label{outage}
\begin{split}
&P_{\text{out}}({\scriptsize{\textsf{SNDR}}},x) = 1 - \frac{2^{\alpha+\beta-2}}{\pi\Gamma(\alpha)\Gamma(\beta)}m{N \choose m}\sum_{n=0}^{m-1}\frac{(-1)^n}{N-m+n+1} \\& {m-1 \choose n} \exp\left(-\frac{(N-m+n+1)\kappa(1+{\scriptsize{\textsf{ILR}}})x}{((N-m+n)(1-\rho)+1)(1-{\scriptsize{\textsf{ILR}}}x)\overline{\gamma}_1}\right)\\&\MeijerG[\Bigg]{5}{0}{0}{5}{-}{\bold{b}}{\frac{(\alpha\beta)^2(\e{\gamma_{1(m)}}+\kappa)(N-m+n+1)x}{16((N-m+n)(1-\rho)+1)(1-{\scriptsize{\textsf{ILR}}}x)\overline{\gamma}_1 \overline{\gamma}_2}}
\end{split}   
\end{equation}
where $\bold{b} = [\frac{\alpha}{2}, \frac{\alpha+1}{2}, \frac{\beta}{2}, \frac{\beta+1}{2}, 0]$.
\begin{proof}
After transforming the exponential into Meijer-G function and applying
the identity \cite[Eq.~(07.34.21.0013.01)]{24}, the outage is derived as (\ref{outage}).
\end{proof}

\subsection{High SNR Analysis}
Using \cite[Eq.~(07.34.06.0001.01)]{24} to expand the Meijer-G function in (\ref{outage}) at high {\scriptsize{\textsf{SNR}}}, the expansion is given by
\begin{equation}
\MeijerG[\Bigg]{5}{0}{0}{5}{-}{\bold{b}}{z} \cong \sum_{k=1}^5 \prod_{j=1, j \neq k}^{5}\Gamma(b_j - b_k)z^{b_k}   
\end{equation}
where $b_k$ is the $k$-th element of the vector $\bold{b}$.

\subsection{Diversity Analysis}
Given that the outage performance saturates at high SNR by the floor caused by the hardware impairments, it is trivial to conclude that the diversity gain $G_d$ is equal to zero. For an ideal hardware and after expanding the Meijer-G function at high {\scriptsize{\textsf{SNR}}}, it can be shown that the diversity gain is given by

\begin{equation}
G_d =\begin{cases}
          \min\left(m,~\frac{\alpha}{2},~\frac{\beta}{2}\right)  \quad &\text{if} \, \rho = 1 \\
          \min\left(1,~\frac{\alpha}{2},~\frac{\beta}{2}\right)  \quad &\text{otherwise} \\
     \end{cases}
\end{equation}

%% file: capacity.tex
\section{Ergodic Capacity}
\subsection{Exact Analysis}
The ergodic capacity, expressed in bits/s/Hz, is defined as the  maximum error-free data transferred by the system channel. It can be written as follows
\begin{equation}\label{defcap}
\mathcal{I}({\scriptsize{\textsf{SNDR}}}) = \mathbb{E}[\log(1 + \varpi {\scriptsize{\textsf{SNDR}}})] 
\end{equation}
where $\varpi=e/2\pi$ for intensity modulation and direct detection (IMDD). The capacity can be calculated by deriving the PDF of the {\scriptsize{\textsf{SNDR}}}. However, an exact closed-form of Eq.~(\ref{defcap}) is not tractable due to the presence of mathematical terms related to the hardware impairments. Thereby, the numerical integration can be performed to evaluate the exact capacity. 
\subsection{Approximation}
In spite of the difficulty to calculate an exact closed-form of the EC, we can derive a simpler expression by referring to the approximation given by \cite[Eq.~(27)]{22}
\begin{equation}\label{approx}
\e{\log\left(1 + \frac{\psi}{\varphi}\right)} \cong \log\left(1+\frac{\e{\psi}}{\e{\varphi}} \right).
\end{equation}
Although Eq.~(\ref{approx}) has no theoretical foundations, it provides with a tight bound on the capacity.
\subsection{Bound I}
For high {\scriptsize{\textsf{SNR}}} values, the {\scriptsize{\textsf{SNDR}}} converges to ${\scriptsize{\textsf{SNDR}}}^*$ defined by
\begin{equation}
\lim_{\overline{\gamma}_1,\overline{\gamma}_2\to\infty}{\scriptsize{\textsf{SNDR}}} = \frac{1}{\frac{(1+{\scriptsize{\textsf{ILR}}})\xi}{\delta}-1} = {\scriptsize{\textsf{SNDR}}}^*. 
\end{equation}
\begin{corollary}
Suppose that $\overline{\gamma}_1$ and $\overline{\gamma}_2$ converge to infinity and the electrical and optical channels are independent, the ergodic capacity converges to a capacity ceiling defined by
\end{corollary}
\begin{equation}
\mathcal{I}({\scriptsize{\textsf{SNDR}}}) \leq \log\left(1+ \varpi {\scriptsize{\textsf{SNDR}}}^*\right). 
\end{equation}

\begin{proof}
Since the {\scriptsize{\textsf{SNDR}}} converges to ${\scriptsize{\textsf{SNDR}}}^*$ as the average {\scriptsize{\textsf{SNR}}}s of the first and second hops largely increase, the Dominated Convergence Theorem allows to move the limit inside the logarithm function.
\end{proof}

\subsection{Bound II}
If the relaying system is linear, i.e, the system is only impaired by IQ imbalance, the {\scriptsize{\textsf{SNDR}}} and the average capacity are saturated at the high {\scriptsize{\textsf{SNR}}} regime as follows
\begin{equation}
 \lim_{ {\scriptsize{\textsf{IBO}}}  \to +\infty}{\scriptsize{\textsf{SNDR}}^*} =  \lim_{ {\scriptsize{\textsf{IBO}}}  \to +\infty} \frac{1}{\frac{(1+{\scriptsize{\textsf{ILR}}})\xi}{\delta}-1} =  \frac{1}{{\scriptsize{\textsf{ILR}}}}.  
\end{equation}
\begin{corollary}
For linear relaying, the system capacity under IQ imbalance is bounded at high {\scriptsize{\textsf{SNR}}} by
\end{corollary}
\begin{equation}
 \mathcal{I}({\scriptsize{\textsf{SNDR}}}) \leq \log\left( 1 + \frac{\varpi}{{\scriptsize{\textsf{ILR}}}} \right).  
\end{equation}

\subsection{Jensen Bound}
To further characterize the ergodic capacity, it is possible to derive the expression of the upper bound stated by the following Theorem.
\begin{theorem}
For asymmetric (Rayleigh/Gamma-Gamma) fading channels, the ergodic capacity under joint effects of HPA nonlinearities and IQ imbalance is bounded by
\end{theorem}
\begin{equation}
    \mathcal{I}({\scriptsize{\textsf{SNDR}}}) \leq \log\left(1 + \frac{\varpi \mathcal{J}}{{\scriptsize{\textsf{ILR}}}\mathcal{J} + 1}\right)
\end{equation}
where $\mathcal{J}$ is given by
\begin{equation}
\begin{split}
 \mathcal{J} = \e{\frac{\gamma_{1(m)}\gamma_{2(m)}}{({\scriptsize{\scriptsize{\textsf{ILR}}}} + (1 + {\scriptsize{\scriptsize{\textsf{ILR}}}})\kappa   )\gamma_{2(m)} + (1+{\scriptsize{\textsf{ILR}}})(\e{\gamma_{1(m)}} + \kappa)}}
 \end{split}
\end{equation}
After some mathematical manipulations, $\mathcal{J}$ is given by
\begin{equation}
\begin{split}
\mathcal{J} =& \ddfrac{m{N \choose m}(\alpha\beta)^\frac{\alpha + \beta}{2}\left( \frac{\e{\gamma_{1(m)}} + \kappa}{\kappa}\right)^\frac{\alpha+\beta}{4}}{2\pi(1+{\scriptsize{\textsf{ILR}}})\kappa\Gamma(\alpha)\Gamma(\beta)\overline{\gamma}_2^\frac{\alpha+\beta}{4}}\\&~
\times \sum_{n=0}^{m-1} {m-1 \choose n}\frac{(-1)^n((N-m+n)(1-\rho)+1)\overline{\gamma}_1}{(N-m+n+1)^2}\\&
\times~\MeijerG[\Bigg]{5}{1}{1}{5}{\lambda_0}{\lambda_1}
{\frac{(\alpha\beta)^2(\e{\gamma_{1(m)}}+\kappa)}{16\kappa\overline{\gamma}_2}}
\end{split}
\end{equation}
where $\lambda_1 = \left[\frac{\alpha - \beta}{4}, \frac{\alpha - \beta + 2}{4}, \frac{\beta - \alpha}{4}, \frac{\beta - \alpha +2 }{4},-\frac{\alpha + \beta}{4}\right]$ and $\lambda_0=-\frac{\alpha + \beta}{4}$. 

%% file: numerical.tex
\section{Numerical results}
\begin{figure}[H]
    \centering
    \includegraphics[width=\linewidth]{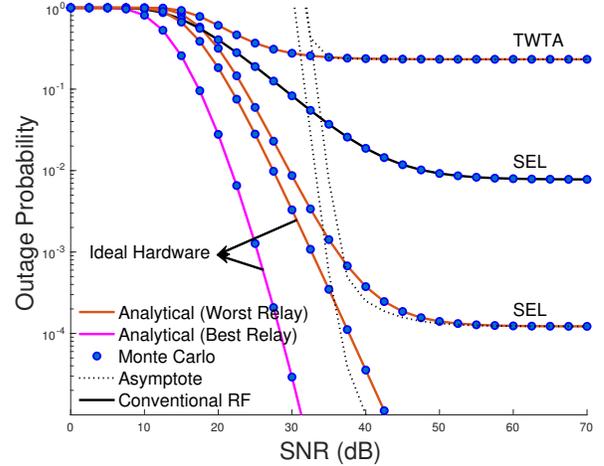}
    \caption{Comparison between the SEL and TWTA and their effects on the outage. The {\scriptsize{\textsf{ILR}}} is fixed at -15 dB while the IBO is fixed at 0 dB.}
    \label{outage3}
\end{figure}

\begin{figure}[H]
    \centering
    \includegraphics[width=\linewidth]{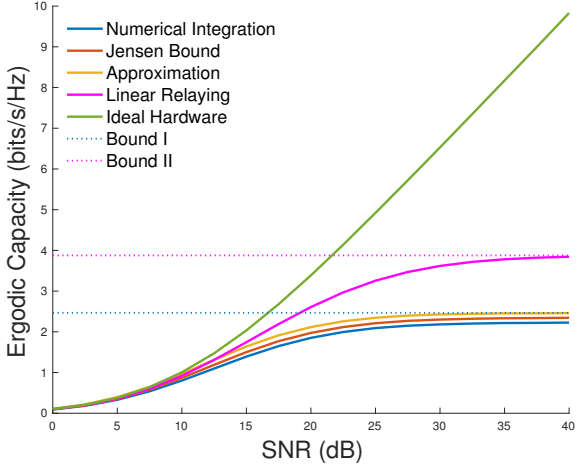}
    \caption{Impacts of the HPA nonlinearities and IQ imbalance on the spectral efficiency. The {\scriptsize{\textsf{ILR}}} is fixed at -15 dB while the IBO is fixed at 0 dB.}
    \label{rate1}
\end{figure}

This Section presents the analytical and numerical\footnote[2]{For all cases, $10^6$ realizations of the random variables were generated to perform the Monte Carlo simulation in MATLAB.} results of the outage probability and ergodic capacity obtained from the mathematical expressions derived in the previous Sections. Unless otherwise stated, we assume that the outage threshold ($x=10$ dB), the temporal correlation ($\rho=0.9$), the number of relay ($N=5$), the rank of selected relay ($m=2$), and the Rytov variance ($\sigma_{R}^2=0.16$).
In Fig.~\ref{outage3}, we observe that the impact of the hardware impairments is not significant at low {\scriptsize{\textsf{SNR}}}, however, this impact becomes remarkable at high {\scriptsize{\textsf{SNR}}} wherein the distortion creates an irreducible outage floor. Besides, we note that the introduction of the wireless optical signaling and multiple relays decreases the outage much lower than the conventional RF relaying systems \cite{22} at low {\scriptsize{\textsf{SNR}}}. At high {\scriptsize{\textsf{SNR}}}, even though both systems are saturated by the outage floor, the proposed system achieves better outage due to the higher diversity order compared to the conventional RF system.
 The same conclusion can be drawn from Fig.~\ref{rate1} wherein the capacity is saturated at high {\scriptsize{\textsf{SNR}}} by a ceiling created by the hardware impairments. We further note that the TWTA introduces more severe deterioration to the system compared to SEL and this result has been confirmed by related works. 

\begin{figure}[H]
    \centering
\includegraphics[width=\linewidth]{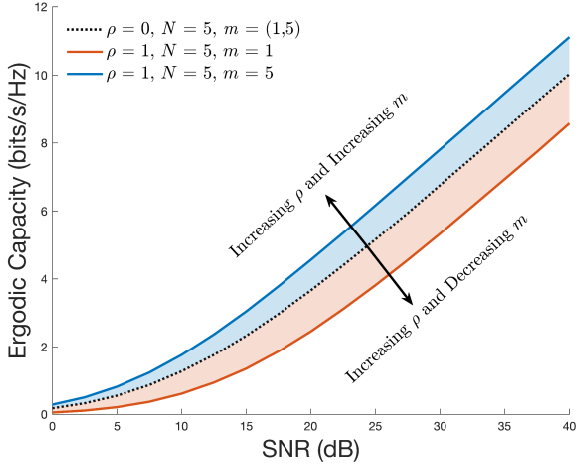}
    \caption{Performance results: Impacts of the temporal correlation and the rank of the selected relay on the spectral efficiency.}
    \label{fig5}
\end{figure}
Fig.~\ref{fig5} provides the dependence of the temporal correlation and the rank of the selected relay on the ergodic capacity. Given that the CSI are sorted in increasing order, then the performance improves with the rank of the selected relay. However, this improvement is achieved only when the CSIs are completely correlated ($\rho \cong 1$), i.e., the CSIs at the moments of relay selection and transmission are roughly the same. In other terms, the channels are slowly varying in time. Inversely, if the channels are uncorrelated ($\rho \cong 0$), i.e., the channels are rapidly varying in time, the source does not have any knowledge about the channels conditions and hence the rank does not reflect the best relay selection, i.e., the higher the rank is, the worse the performances are.

%% file: conclusion.tex
\section{Conclusion}
In this work, we provided the analysis of various models of impairments and their effects on the system performance. We introduced the SEL and TWTA as HPA nonlinearities affecting the relays and we assume that the destination is impaired by IQ imbalance. Moreover, we concluded that the TWTA has more severe impact on the system performance than the SEL model. Furthermore, even though the performance deteriorates under the effects of the imperfections, we noted that the introduction of the FSO technique and multiple relays makes the mixed RF/FSO system more resilient to the hardware impairments than the conventional RF relaying systems. Besides, we illustrated the performance degradation by the hardware impairments as a form of irreducible outage floor as well as the capacity ceiling at high {\scriptsize{\textsf{SNR}}}. Finally, we investigated the impact of the temporal CSI correlation and we concluded that the performance improvement is conditioned upon the channels correlation.